\documentclass[runningheads]{llncs}
\usepackage[T1]{fontenc}
\usepackage{graphicx}
\usepackage{amsmath,amsfonts,amssymb}
\usepackage{booktabs}
\usepackage{array}
\usepackage{multirow}
\usepackage{siunitx}
\usepackage{subcaption}
\usepackage{xcolor}
\usepackage{threeparttable}
\usepackage{makecell}
\usepackage{bbm}
\usepackage{microtype}
\usepackage{tabularx}
\usepackage[version=4]{mhchem}
\usepackage{enumitem}

\usepackage{color}
\usepackage{hyperref}

\newenvironment{tight_enumerate}{
\begin{enumerate}
  \setlength{\itemsep}{0pt}
  \setlength{\parskip}{0pt}
}{\end{enumerate}}

\begin{document}
\title{Personalized AI Practice Replicates Learning Rate Regularity at Scale}
\titlerunning{Personalized AI Practice Replicates Learning Rate Regularity}
\author{Jocelyn Beauchesne\inst{1} \and
Christine Maroti\inst{1} \and
Jeshua Bratman\inst{1} \and
Jerome Pesenti\inst{1} \and
Laurence Holt\inst{1} \and
Alex Tambellini\inst{1} \and
Allison McGrath\inst{1} \and
Matthew Guo\inst{1} \and
Sarah Peterson\inst{1}}
\authorrunning{J. Beauchesne et al.}
\institute{Campus AI\\
\email{contact@campus.ai}}
\maketitle
\begin{abstract}
Recent research \cite{koedinger2023astonishing} demonstrated that students exhibit consistent learning rates across diverse educational contexts. We test these findings using a dataset of 1.8 million (366k post-filtering) student interactions from the digital platform Campus AI providing further evidence to the observation of regularity in learning rate among students. Unlike prior work requiring manual cognitive modeling, Campus AI automatically generates Knowledge Components (KCs) \cite{koedinger2012kli} and corresponding exercises, both of which are validated by human experts. This one-to-many mapping facilitates the application of Additive Factors Models to measure learning parameters without complex cognitive modeling.

Using mixed-effects logistic regression, we confirmed the core finding of prior work \cite{koedinger2023astonishing,simpson2024replicating}: students displayed substantial variation in initial knowledge ($\text{IQR} = [2.78, 12.18]$ practice opportunities to reach 80\% mastery) but remarkably consistent learning rates ($\text{IQR} = [7.01, 8.25]$ opportunities). Furthermore, students using this fully automated system achieved 80\% mastery in a median of 7.22 practice opportunities, comparable to the 6.54 reported for expert-designed curricula \cite{koedinger2023astonishing}. These results suggest that automated, science-grounded content generation can support effective personalized learning at scale. Data and code are publicly available.\footnote{\url{https://github.com/Campus-edu-AI/learning-rate}}

\end{abstract}
\section{Introduction}

Modeling student learning trajectories is essential for understanding cognition and advancing educational technology. However, a persistent bottleneck in the development of personalized learning systems is the "content mapping problem." Traditional systems require domain experts to manually decompose curricula into Knowledge Components (KCs) and map them to specific exercises which is a labor-intensive process that limits scalability.

Recent advances in Large Language Models (LLMs) offer a potential solution by automating the generation of KCs and exercises. Yet, a critical question remains: \textit{Does automated, AI-generated content support the same quality of learning as expert-designed curricula?}

We address this question by replicating the findings of \cite{koedinger2023astonishing}, who identified an "astonishing regularity" in student learning: despite vast differences in initial knowledge, students improve at remarkably consistent rates per practice opportunity on well-defined skills. A practice opportunity under favorable learning conditions, as defined in \cite{koedinger2023astonishing}, consists of attempting a KC-targeted task with immediate feedback, scaffolded hints on demand, the opportunity to retry until correct, and varied practice across multiple exercises. We hypothesize that if AI-generated KCs are pedagogically valid, they should elicit this same fundamental cognitive signature.

Using interaction data from Campus AI, a platform that utilizes LLMs to generate KCs and exercises directly from student materials, we analyze over 360,000 practice opportunities validated by expert review. This system circumvents manual tagging by creating a fully automated, one-to-many mapping between KCs and exercises.

Adopting the analytical framework of \cite{koedinger2023astonishing}, we confirmed that the learning rate regularity persists even with fully AI-generated content. Furthermore, students using Campus AI's automated system reached 80\% mastery in a median of 7.22 practice opportunities (comparable to the 6.54 reported for expert-authored systems). These results suggest that Generative AI can effectively bypass the content bottleneck, supporting personalized learning at scale without compromising pedagogical validity. The dataset, examples of generated KCs and exercises, and a fully replicable analysis pipeline are available at \url{https://github.com/Campus-edu-AI/learning-rate}.

\section{Background and Related Work}
\subsection{The Knowledge Learning Instruction Framework}

The Knowledge-Learning-Instruction (KLI) framework \cite{koedinger2012kli} decomposes learning objectives into trackable units called Knowledge Components (KCs). A KC represents a distinct, actionable piece of knowledge—an unobservable cognitive process characterized by specific \textit{conditions} and a resulting \textit{response}.

KCs operate through a condition-response mechanism mirroring expert performance:

\begin{tight_enumerate}
    \item \textbf{Conditions of Application}: Cues signaling when to apply knowledge. These may be \textit{constant} (specific symbols) or \textit{variable} (categories sharing features).
    \item \textbf{Response}: The action produced. Responses may be \textit{constant} (fixed outputs) or \textit{variable} (adaptive to context).
\end{tight_enumerate}

Table~\ref{tab:kc-examples} illustrates examples of generated KCs across different types and subjects; additional examples are available in our repository.\footnote{\url{https://github.com/Campus-edu-AI/learning-rate}}

\begin{table}[h]
\centering
\small
\begin{tabular}{lp{4cm}p{6.2cm}}
\toprule
\textbf{Type} & \textbf{Condition} & \textbf{Response} \\
\midrule
Category & Random variable types & Discrete or continuous \\
Fact & U wave on EKG & Late repolarization of Purkinje fibers \\
Rule/Plan & Eye focus during basketball shot & On the target (front or back of rim) \\
Concept & Contractual capacity & Legal ability to enter into a binding contract \\
\bottomrule
\end{tabular}
\caption{Examples of LLM-generated Knowledge Components across different types.}
\label{tab:kc-examples}
\end{table}

\subsection{Learning Rate Regularity}

We build upon the "astonishing regularity" in learning rates identified by \cite{koedinger2023astonishing}. Using the individualized Additive Factors Model (iAFM) \cite{cen2006learningfactors,liu2017towards} across 1.3 million interactions, they found that while initial knowledge varies, improvement rates per opportunity remain consistent (SD = 0.015 log odds), with students averaging 7 opportunities to reach 80\% mastery. \cite{simpson2024replicating} replicated this using Carnegie Learning's MATHia system, confirming consistency across platforms. Critically, these studies relied on expert-designed KCs. We investigate whether this regularity persists when KCs and exercises are fully LLM-generated. A positive result would suggest learning rate consistency reflects fundamental cognitive properties rather than instructional design artifacts.

\subsection{Automated Educational Content Generation}

Recent work demonstrates that Large Language Models (LLMs) can generate pedagogically viable educational content. \cite{sarsa2022automatic} found LLM-generated programming exercises to be largely novel and usable, while \cite{xiao2023evaluating} showed ChatGPT-generated reading comprehension matched or exceeded human textbook quality. Similarly, \cite{olney2023generating} reported that LLM-generated multiple-choice questions rivaled human-authored content on most quality metrics. Further research confirms the utility of fine-tuned models for topic-specific questions \cite{li2025novel,noorbakhsh2025savaal} and automated KC tagging \cite{moore2024automated}. While these studies rely primarily on expert judgment for validation, we propose a complementary approach: validating AI-generated content by replicating the learning rate regularities observed in expert-designed systems, thereby demonstrating measurable learning outcomes.

\section{Campus AI System Overview}

Campus AI (formerly called Sizzle AI) is a mobile platform designed for personalized test preparation through scaffolded, bite-size-learning practice. The system offers varied practice, immediate feedback, and explanatory hints. Users initiate learning by uploading documents (such as readings, study guides, practice tests etc) or selecting topics, prompting the automated generation of Knowledge Components (KCs),and  Exercises (Multiple Choice, Fill-In-The-Blank, Pair Matching, Highlight-The-Mistake) which are intelligently provided as practice opportunities through an intelligent exercise sequencing system.
\subsection{Knowledge Component Generation}

Using the student's input, the system employs LLMs to extract KCs following the KLI framework \cite{koedinger2012kli}. Each KC consists of a \textit{condition} triggering application and an expected \textit{response}. KCs are organized hierarchically into skills and units.

To validate KC quality, four college instructors evaluated 361 generated KCs against their course materials across three dimensions: (1) \textbf{Clarity}—whether the KC is specific, unambiguous, and assessable; (2) \textbf{Relevance}—how important mastery is for achieving the learning objective; and (3) \textbf{Difficulty}—whether the KC is appropriately calibrated for the course level.

\begin{table}[h]
\centering
\begin{minipage}{0.31\textwidth}
\centering
\begin{tabular}{lr}
\toprule
\textbf{Clarity} & \textbf{\%} \\
\midrule
Very clear     & 75.3 \\
Somewhat clear & 12.7 \\
Not clear      & 11.9 \\
\bottomrule
\end{tabular}
\end{minipage}%
\hfill
\begin{minipage}{0.31\textwidth}
\centering
\begin{tabular}{lr}
\toprule
\textbf{Relevance} & \textbf{\%} \\
\midrule
Very important     & 60.7 \\
Somewhat important & 15.8 \\
Not important      & 23.0 \\
\bottomrule
\end{tabular}
\end{minipage}%
\hfill
\begin{minipage}{0.31\textwidth}
\centering
\begin{tabular}{lr}
\toprule
\textbf{Difficulty} & \textbf{\%} \\
\midrule
Just right   & 86.5 \\
Too basic    & 8.2  \\
Too advanced & 5.3  \\
\bottomrule
\end{tabular}
\end{minipage}
\caption{KC evaluation by instructors: clarity (n=361), relevance (n=359), difficulty (n=282).}
\label{tab:kc-eval}
\end{table}
Instructors rated 88\% of KCs as clear, 77\% as relevant to learning objectives, and 87\% as appropriately calibrated for difficulty.

\subsection{Exercise Generation}
For each KC, an LLM-based pipeline generates diverse exercises. To ensure quality, we employ an "LLM-as-a-Judge" approach \cite{zheng2023judgingllmasajudgemtbenchchatbot,bavaresco2025llmsinsteadhumanjudges}, utilizing a panel of models to filter content \cite{verga2024replacingjudgesjuriesevaluating}. In production, this pipeline rejects approximately 36\% of candidates.

We further validated post-filtering exercise quality through an expert review by eight AP teachers across multiple subject areas, who rated 446 exercises on a five-point scale: ``Good,'' ``Incorrect,'' ``Irrelevant,'' ``Inappropriate,'' or ``Confusing.'' The teachers have 7--34 years of experience (16 on average) and teach at a mix of public, private, and charter schools.

Of the 446 exercises reviewed, 86.5\% were rated ``Good,'' while only 2.2\% were labeled factually incorrect; the remainder were flagged as confusing (5.6\%), irrelevant (2.0\%), or having multiple valid answers (3.6\%). Qualitative feedback highlighted two areas for improvement: (1) difficulty calibration for AP-level students, and (2) emphasis on conceptual understanding over specific details. Due to time and resource constraints, only a subset of 37 exercises were labeled by more than one teacher. Inter-annotator agreement on this subset yielded Cohen's $\kappa$ of 0.54: moderate agreement consistent with known variability in human quality judgments. The two teachers agreed that 26 out of 37 (70\%) of exercises in this subset were ``Good.''

\subsection{Exercise Sequencing and Learning Conditions}
Campus AI implements a quiz-first learning approach in which students practice a selected skill through a "feed" made up of a sequence of exercises. Critically, the sequencing algorithm does not meaningfully rely on student performance, avoiding bias toward KCs the student has mastered or away from those they struggle with. Instead, it randomly samples Knowledge Components to prioritize those with fewer practice opportunities. With enough attempts, this converges towards a uniform practice across all Knowledge Components within a skill.

To improve student momentum and engagement, the algorithm initially presents simplified versions of exercises, for example one distractor instead of 3. This helps the student build momentum and increases the number of total exercises attempted in one session. Note, however, this detail also introduces a bias: the first two exercises for each Knowledge Component are easier and thus have a higher chance of being answered correctly. We account for this in our analysis by including the exercise type and simplification indicator in our model as fixed effects.

\section{Dataset}

For this study, we collected 1.8 million practice opportunities. Each record represents a student interaction with an exercise mapped to a KC $\kappa$, containing a timestamp, first-attempt correctness label, identifiers (student, KC, exercise), metadata (type, subject, level), and the opportunity count $T_{s, \kappa}$.

Campus AI generates Knowledge Components uniquely for each student's uploaded materials, so most Knowledge Components have interactions with only one student. This contrasts with \cite{koedinger2023astonishing}, where at least 10 students interacted with at least 2 practice opportunities for the same Knowledge Component.

Our dataset, before filtering, has 567k unique Knowledge Components across 13k unique students. Because KCs are generated independently from each student's uploaded materials, students studying similar topics from different sources produce distinct KCs that resist deduplication even with embedding-based similarity approaches.

\subsection{Filtering}
To ensure robust observation of learning, we exclude KCs with fewer than 5 interactions, as learning trends cannot be reliably estimated from sparse data. Additionally, we exclude practice opportunities beyond the 30th to mitigate outlier effects. This filtering removes the long tail of brief user engagements. The final dataset comprises 366,127 practice opportunities across 51,437 unique KCs and 7,161 students. Each student practiced a median of 3 KCs (IQR: 1--7).

\section{Cognitive Modeling}
Following \cite{koedinger2023astonishing}, we modeled the probability of a correct response using a mixed-effects logistic regression (iAFM variant). Our model diverges in two respects:
\begin{tight_enumerate}
    \item We exclude KC-specific terms, as KCs are rarely shared across students.
    \item We include fixed effects for exercise type and simplification status to account for varying baseline difficulties.
\end{tight_enumerate}

Let $p_{s, e}$ denote the probability that student $s$ correctly answers exercise $e$ (testing KC $\kappa$ at step $T$) on the first attempt:

\begin{equation}
\label{eq:base_model}
\log\left(\frac{p_{s, e}}{1 - p_{s, e}}\right) = \theta + \theta_s + (\delta + \delta_s) \cdot T_{s, \kappa} + \beta_{\text{type}(e)} + \gamma S_e
\end{equation}

Here, $\theta$ and $\theta_s$ represent population and student-specific intercepts (prior knowledge); $\delta$ and $\delta_s$ denote the global learning rate and student-specific deviation; and $T_{s, \kappa}$ is the count of prior practice opportunities. Additionally, $\beta_{\text{type}(e)}$ captures exercise type difficulty, while $\gamma$ is the coefficient for the simplification indicator $S_e$.

To facilitate convergence, $T_{s, \kappa}$ was scaled by $0.01$, with learning rate effects rescaled accordingly.

\section{Results}

Consistent with \cite{koedinger2023astonishing}, we observed a significantly wider variation in initial knowledge than in learning rates, that is, students may vary in starting point, but the speed at which they learn is consistent across the population. The interquartile range (IQR) for initial knowledge was 15.32\%, compared to 0.26\% for learning rates. Additionally, the standard deviation of student learning rates was 0.0122, aligning closely with the 0.015 reported by \cite{koedinger2023astonishing}.

\begin{table}[!htbp]
    \centering
    \caption{Summary Statistics for Mixed-Effects Model Parameters}
    \label{tab:model-parameters}
    \begin{tabular}{@{}lcccccc@{}}
        \toprule
        \textbf{Parameter} & \textbf{Population} & \multicolumn{5}{c}{\textbf{Student Effect}} \\
        \cmidrule(lr){3-7}
        & \textbf{Effect} & \textbf{Mean} & \textbf{Q1} & \textbf{Median} & \textbf{Q3} & \textbf{SD} \\
        \midrule
        Initial Knowledge & 0.686 & -0.0169 & -0.304 & 0.0221 & 0.314 & 0.461 \\
        Learning Rate     & 0.0657 & -0.000262 & -0.00555 & 0.000323 & 0.00516 & 0.0121 \\
        \bottomrule
    \end{tabular}
    \begin{tablenotes}
        \small
        \item Population effects represent the fixed effects from the mixed-effects logistic regression model. Individual-level statistics show the distribution of random effects across n=7161 students. Initial knowledge values are in log odds, learning rates are in log odds per practice opportunity.
    \end{tablenotes}
\end{table}

\begin{figure}[!htbp]
    \centering
    \includegraphics[width=0.7\linewidth]{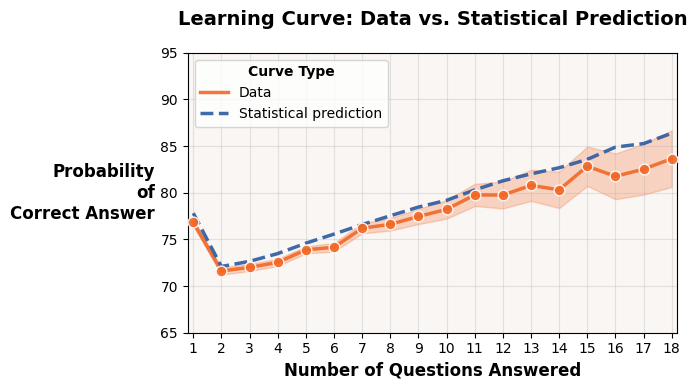}
    \caption{Observed learning curve versus statistical model predictions across practice opportunities on a Knowledge Component. The solid orange line shows empirical accuracy rates from student practice data. The dashed blue line represents the population-level prediction from a mixed-effects logistic regression model (iAFM framework). Students show consistent improvement from approximately 73\% to 85\% accuracy over 18 practice opportunities.}
    \label{fig:learning-rate}
\end{figure}

\begin{figure}[!htbp]
    \centering
    \begin{subfigure}[b]{0.49\textwidth}
        \centering
        \includegraphics[width=\textwidth, height=6cm, keepaspectratio=false]{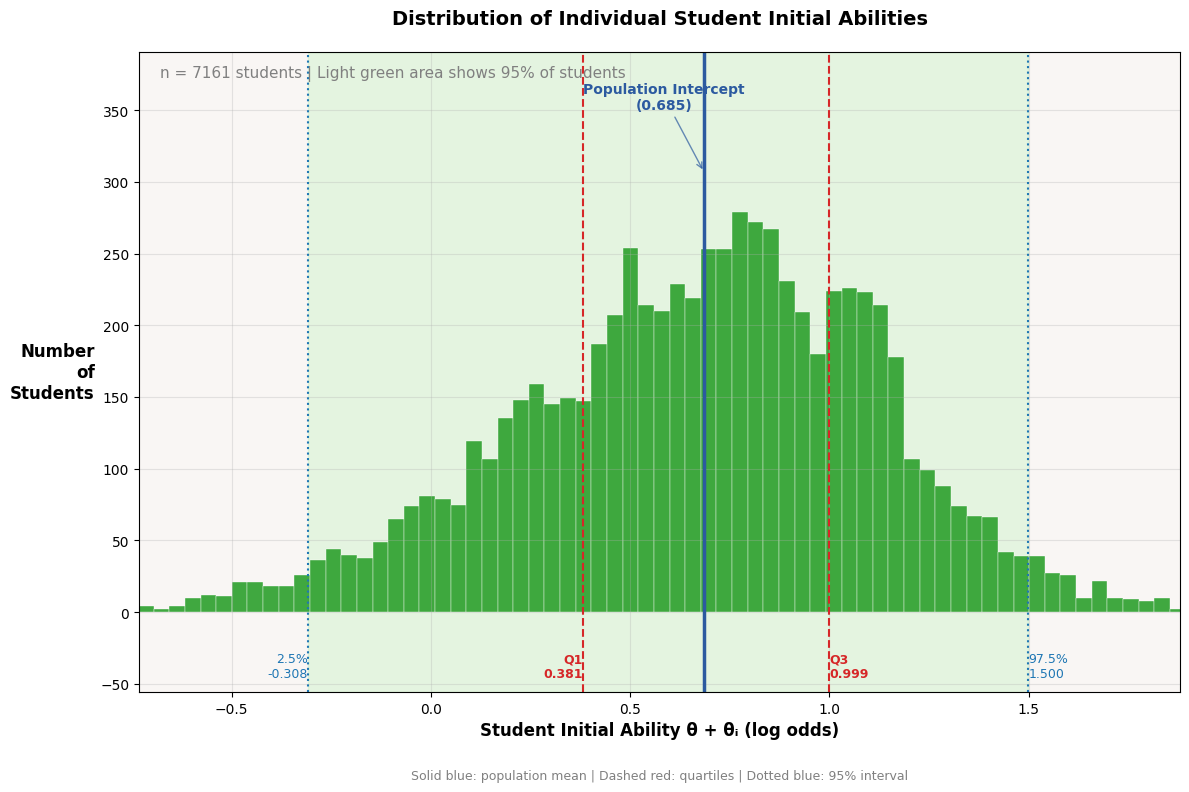}
        \caption{Distribution of individual student initial abilities ($\theta + \theta_s$) in log odds. The population intercept is 0.685 with 95\% of students falling within the light green shaded region. The distribution shows substantial individual variation around the population mean.}
        \label{fig:base-model-theta}
    \end{subfigure}
    \hfill
    \begin{subfigure}[b]{0.49\textwidth}
        \centering
        \includegraphics[width=\textwidth, height=6cm, keepaspectratio=false]{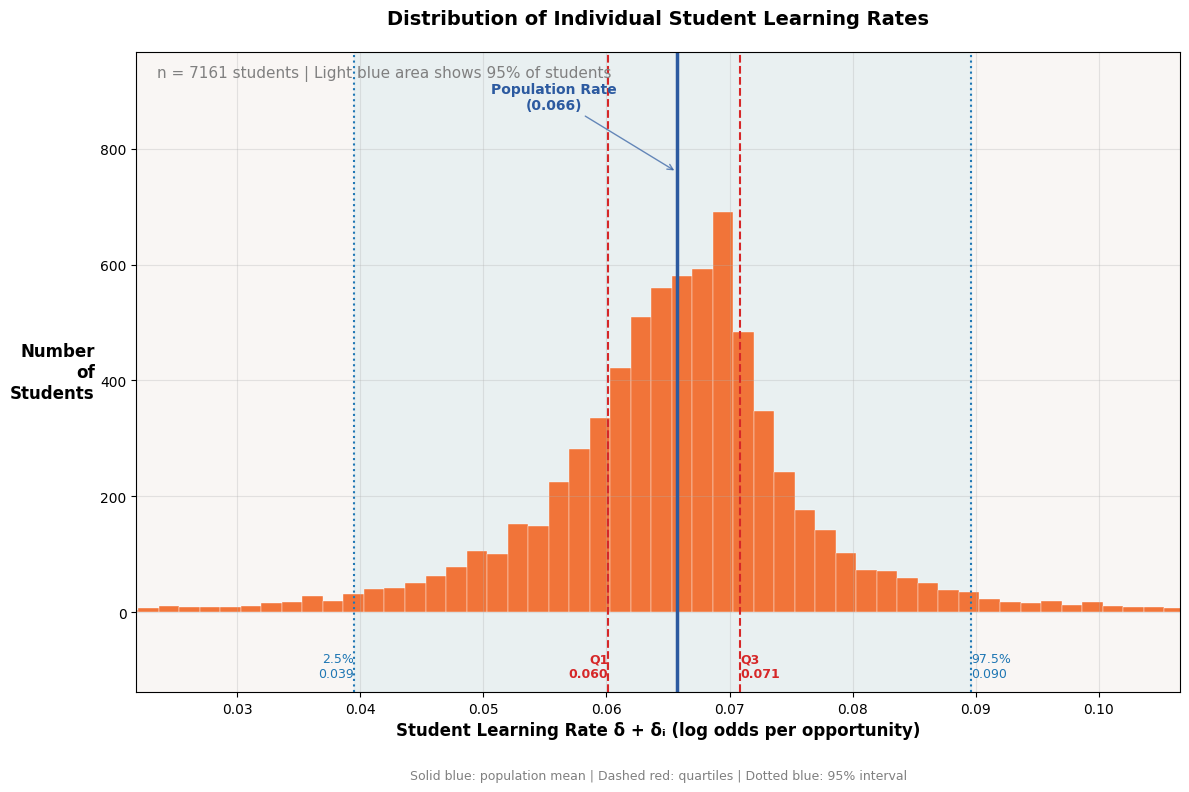}
        \caption{Distribution of individual student learning rates ($\delta + \delta_s$) in log odds per opportunity. The population rate is 0.066 with most students concentrated near the population mean, showing relatively consistent learning rates across individuals.}
        \label{fig:base-model-delta}
    \end{subfigure}
    \caption{Parameter distributions from the base mixed-effects logistic regression model (iAFM framework) for n=7161 students. Both distributions show the individual-level variation around population parameters, with initial abilities displaying greater heterogeneity than learning rates. Quartile boundaries (Q1, Q3) and 95\% confidence intervals are marked for reference.}
    \label{fig:base-model-distributions}
\end{figure}

\begin{table}[!htbp]
\caption{Median for initial accuracy and practice opportunities to reach mastery ($80\%$ correct probability) for low (25) and high (75) percentiles of initial knowledge (assuming overall learning rate, $\delta$) and learning rate (assuming overall initial knowledge, $\theta$) for the base models.}
\label{tab:knowledge-learning-rate}

\centering
\sisetup{table-align-text-post=false}
\renewcommand{\theadfont}{\normalsize\bfseries}
\begin{tabular}{
    c
    S[table-format=1.2]
    S[table-format=2.2]
    S[table-format=2.2]
    S[table-format=2.2]
    S[table-format=1.2]
    S[table-format=2.2]
}
\toprule
& \multicolumn{3}{c}{\bfseries Initial Knowledge} & \multicolumn{3}{c}{\bfseries Learning Rate} \\
\cmidrule(lr){2-4} \cmidrule(lr){5-7}
\thead{Percentile} & {\thead{Log Odds}} & {\thead{Percent \\ Correct}} & {\thead{Ops to 80\% \\ (Fixed Rate)}} & {\thead{Log Odds \\ per Opp}} & {\thead{Percent Point \\ Improvement}} & {\thead{Ops to 80\% \\ (Fixed Knowledge)}} \\
\midrule
25 & 0.59 & 64.24 & 12.18 & 0.06 & 1.23 & 8.25 \\
50 & 0.91 & 71.34 & 7.22  & 0.07 & 1.34 & 7.52 \\
75 & 1.20 & 76.92 & 2.78  & 0.07 & 1.44 & 7.01 \\
\bottomrule
\end{tabular}
\end{table}

\section{Ablation study}

\subsection{Motivation}
For each pair of student $s$ and Knowledge Component $\kappa$, we have additional information extracted by an LLM, namely the Course Subject (Math, History, etc.), Course Level (High School, College Senior, etc.), and Knowledge Component Type (Fact, Procedure, etc.). The motivation is to explore questions such as:
\begin{tight_enumerate}
    \item Are certain Knowledge Component Types "easier" to learn than others? For example, Facts versus Procedures.
    \item Are some topics "harder" to learn than others? For example, Math versus History.
    \item Does the level of a student explain the speed at which they learn? For example Middle School versus College Senior.
\end{tight_enumerate}

\subsection{Extended model}

We extend our model Equation \eqref{eq:base_model}:
\begin{align}
\log\left(\frac{p_{s, e}}{1 - p_{s, e}}\right) = \,&\theta + \theta_s + \mathbbm{1}_{\text{subject}} \cdot \theta_{\text{subject}} + \mathbbm{1}_{\text{level}} \cdot \theta_{\text{level}} + \mathbbm{1}_{\kappa_{\text{type}}} \cdot \theta_{\kappa_{\text{type}}} \nonumber \\
&+ (\delta + \delta_s + \mathbbm{1}_{\text{subject}} \cdot \delta_{\text{subject}} + \mathbbm{1}_{\text{level}} \cdot \delta_{\text{level}} + \mathbbm{1}_{\kappa_{\text{type}}} \cdot \delta_{\kappa_{\text{type}}}) \cdot T_{s, \kappa} \nonumber \\
&+ \beta_{e} + \gamma S_e
\end{align}

where $\mathbbm{1}_{\text{subject}}, \mathbbm{1}_{\text{level}}, \mathbbm{1}_{\kappa_{\text{type}}} \in \{0, 1\}$ indicate whether each factor is included in the model; $\theta_{\text{subject}}, \theta_{\text{level}}, \theta_{\kappa_{\text{type}}}$ are fixed effects on prior knowledge for course subject, course level, and KC type respectively; $\delta_{\text{subject}}, \delta_{\text{level}}, \delta_{\kappa_{\text{type}}}$ are fixed effects on learning rate for course subject, course level, and KC type respectively.

\subsection{Results}

We fitted a regression for every possible combination of factors, for a total of 8 models (including the baseline).

\subsubsection{Overall Model Comparison}

Table \ref{tab:ablation_population_student} shows that population-level parameters $\theta_{\text{pop}}$ and $\delta_{\text{pop}}$ remain largely consistent across models (ranging from 0.663 to 0.694 for $\theta_{\text{pop}}$ and 0.064 to 0.068 for $\delta_{\text{pop}}$), suggesting these effects estimates are robust.

Notably, the expected value of student-specific initial knowledge effects ($\mathbb{E}[\theta_s]$) decreases substantially when these group-level fixed effects are added, dropping from -0.017 in the base model to -0.001 in the full model (Model 7). This order of magnitude reduction suggests that the course level, subject, and KC type effects capture variations previously attributed to individual students' initial knowledge.

However, the standard deviation of students' initial knowledge $\theta_s$ remains stable around 0.45-0.46, indicating that most individual differences persist even after accounting for these effects. Similarly, students' learning rates show minimal change across models. The standard deviation of $\delta_s$ remains consistently around 0.012, and the mean stays near zero across all models. This confirms that learning rates are consistent across students, regardless of course level, subject, or KC type. Furthermore, the grouping factor effects ($\theta$ and $\delta$ for KC type, course level, and course subject) remain stable across model variants, with values staying within the same order of magnitude despite slight variations.

\begin{table}[htbp]
\caption{Results of ablation comparing population-level and student-level parameters across model variants. Each model includes different combinations of course level, subject, and KC type effects. Population effects ($\theta_{\text{pop}}$, $\delta_{\text{pop}}$) represent fixed intercept and slope, while student-level statistics show the distribution of random effects across n=7161 students. Initial knowledge values are in log odds; learning rates are in log odds per practice opportunity.}
\label{tab:ablation_population_student}
\begin{tabular}{lccccccccc}
\toprule
    & \text{level} & \text{subject} & \text{kc type} & $\theta_{\text{pop}}$ & $\delta_{\text{pop}}$ & $\mathbb{E}[\theta_{\text{s}}]$ & $\text{SD}[\theta_{\text{s}}]$ & $\mathbb{E}[\delta_{\text{s}}]$ & $\text{SD}[\delta_{\text{s}}]$ \\
\midrule
model 0 &   &   &   & $6.85 e^{\text{-01}}$ & $6.57 e^{\text{-02}}$ & $-1.69 e^{\text{-02}}$ & $4.61 e^{\text{-01}}$ & $-3.00 e^{\text{-04}}$ & $1.21 e^{\text{-02}}$ \\
model 1 & \checkmark &   &   & $6.94 e^{\text{-01}}$ & $6.82 e^{\text{-02}}$ & $-6.30 e^{\text{-03}}$ & $4.60 e^{\text{-01}}$ & $0.00 e^{\text{+00}}$ & $1.20 e^{\text{-02}}$ \\
model 2 &   & \checkmark &   & $6.86 e^{\text{-01}}$ & $6.56 e^{\text{-02}}$ & $-3.37 e^{\text{-03}}$ & $4.55 e^{\text{-01}}$ & $-1.00 e^{\text{-04}}$ & $1.20 e^{\text{-02}}$ \\
model 3 &   &   & \checkmark & $6.63 e^{\text{-01}}$ & $6.42 e^{\text{-02}}$ & $-1.97 e^{\text{-03}}$ & $4.59 e^{\text{-01}}$ & $-0.00 e^{\text{+00}}$ & $1.19 e^{\text{-02}}$ \\
model 4 & \checkmark & \checkmark &   & $6.90 e^{\text{-01}}$ & $6.64 e^{\text{-02}}$ & $-1.93 e^{\text{-03}}$ & $4.55 e^{\text{-01}}$ & $0.00 e^{\text{+00}}$ & $1.19 e^{\text{-02}}$ \\
model 5 & \checkmark &   & \checkmark & $6.86 e^{\text{-01}}$ & $6.63 e^{\text{-02}}$ & $-1.41 e^{\text{-03}}$ & $4.59 e^{\text{-01}}$ & $0.00 e^{\text{+00}}$ & $1.19 e^{\text{-02}}$ \\
model 6 &   & \checkmark & \checkmark & $6.66 e^{\text{-01}}$ & $6.42 e^{\text{-02}}$ & $-1.13 e^{\text{-03}}$ & $4.54 e^{\text{-01}}$ & $-0.00 e^{\text{+00}}$ & $1.19 e^{\text{-02}}$ \\
model 7 & \checkmark & \checkmark & \checkmark & $6.73 e^{\text{-01}}$ & $6.52 e^{\text{-02}}$ & $-1.06 e^{\text{-03}}$ & $4.54 e^{\text{-01}}$ & $0.00 e^{\text{+00}}$ & $1.18 e^{\text{-02}}$ \\
\bottomrule
\end{tabular}
\end{table}

\subsubsection{Exercise Type Effects}

Exercise type effects ($\beta$) remained stable across all model variants, with coefficients reflecting the intuitive difficulty ordering, where $>$ means harder than: $\text{Pair Matching} > \text{Fill-In-The-Blank} > \text{Multiple Choice} > \text{Highlight-The-Mistake}$.

Pair Matching and Fill-In-The-Blank are the hardest exercise types because they are compounded questions. Highlight-The-Mistake is the easiest exercise type because it allows multiple attempts.

\subsubsection{KC Type Effects}

Table \ref{tab:kc_type_effects} shows that students typically demonstrate higher initial knowledge and faster learning rates for declarative KC types (association, category, concept, fact) compared to procedural KC types (principle/rule/model, production/schema/skill, rule/plan).

Declarative types show positive effects $\theta \in [0.014, 0.072]$ and $\delta \in [0.0009, 0.0045]$, while procedural types show negative effects $\theta \in [-0.067, -0.092]$ and $\delta \in [-0.0042, -0.0058]$.

This aligns with the intuition that procedural KCs require linking multiple knowledge pieces together, making them harder to learn \cite{koedinger2012kli}.

\begin{table}[htbp]
\caption{Comparing Intercepts and Learning Rates for the KC Types factor, expected value and standard deviation are computed over results from model 3, 5, 6 and 7.}
\label{tab:kc_type_effects}
\centering
\begin{tabular}{lcccc}
\toprule
    & $\mathbb{E}[\theta_{\text{kc type}}]$ & $\mathbb{E}[\delta_{\text{kc type}}]$ & $\text{SD}[\theta_{\text{kc type}}]$ & $\text{SD}[\delta_{\text{kc type}}]$ \\
KC type &  &  &  &  \\
\midrule
association & $1.42 e^{\text{-02}}$ & $8.95 e^{\text{-04}}$ & $1.76 e^{\text{-03}}$ & $1.32 e^{\text{-04}}$ \\
category & $7.16 e^{\text{-02}}$ & $4.53 e^{\text{-03}}$ & $5.11 e^{\text{-03}}$ & $4.25 e^{\text{-04}}$ \\
concept & $1.83 e^{\text{-02}}$ & $1.16 e^{\text{-03}}$ & $5.39 e^{\text{-03}}$ & $3.67 e^{\text{-04}}$ \\
fact & $2.80 e^{\text{-02}}$ & $1.77 e^{\text{-03}}$ & $3.52 e^{\text{-03}}$ & $2.62 e^{\text{-04}}$ \\
principle/rule/model & $-6.89 e^{\text{-02}}$ & $-4.34 e^{\text{-03}}$ & $6.52 e^{\text{-03}}$ & $3.30 e^{\text{-04}}$ \\
production/schema/skill & $-6.65 e^{\text{-02}}$ & $-4.19 e^{\text{-03}}$ & $1.56 e^{\text{-03}}$ & $1.77 e^{\text{-04}}$ \\
rule/plan & $-9.23 e^{\text{-02}}$ & $-5.82 e^{\text{-03}}$ & $3.45 e^{\text{-03}}$ & $2.02 e^{\text{-04}}$ \\
\bottomrule
\end{tabular}
\end{table}

\subsubsection{Course Subject Effects}

\begin{figure}[!htbp]
    \centering
    \includegraphics[width=0.95\linewidth]{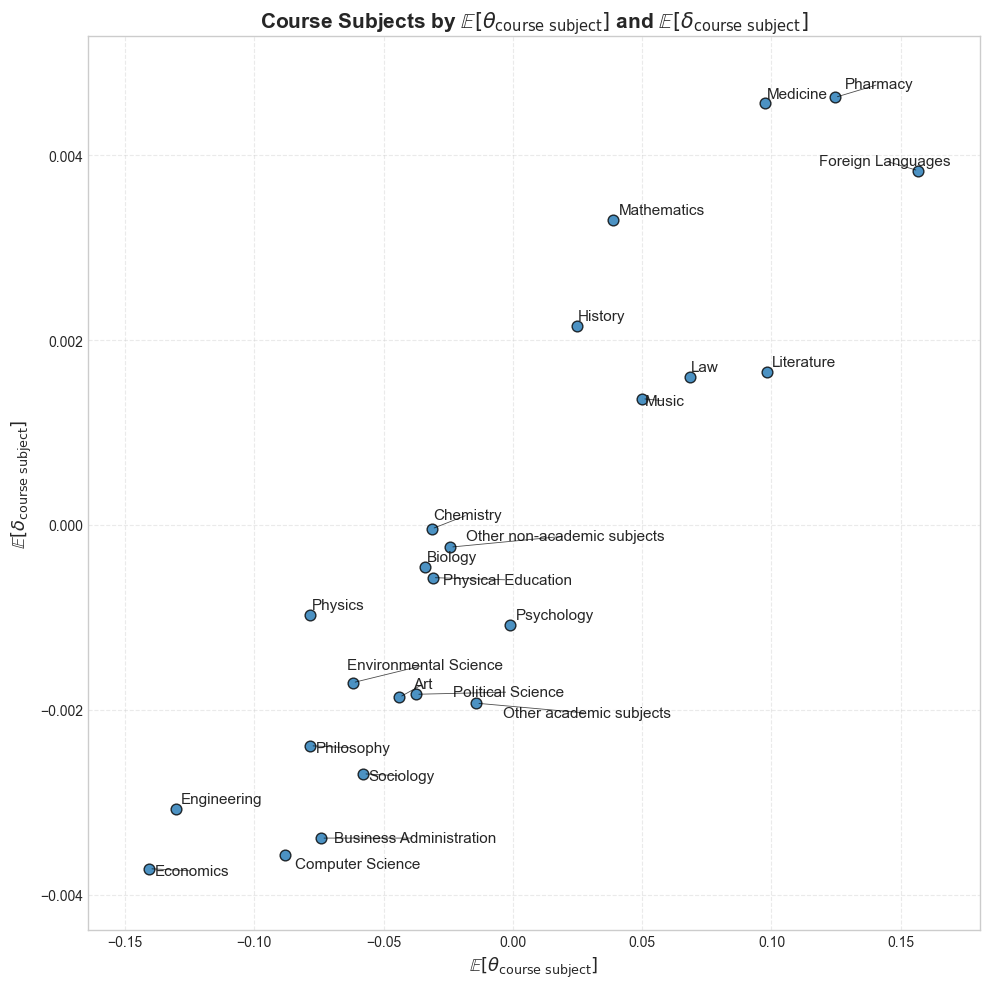}
    \caption{Scatter plot the course subject factor effects, Average $\theta_{\text{course subject}}$ against $\delta_{\text{course subject}}$ averaged across ablation Models 2, 4, 6, and 7.}
    \label{fig:course-subject-scatter-plot}
\end{figure}

Figure \ref{fig:course-subject-scatter-plot} plots course subject effects for initial knowledge ($\theta_{\text{course subject}}$) against learning rate ($\delta_{\text{course subject}}$), highlighting a similar declarative-procedural dichotomy.

Subjects emphasizing memorization and factual knowledge (Medicine, Pharmacy and Foreign Languages) cluster in the upper right corner with positive effects ($\theta \in [0.097, 0.156]$, $\delta \in [0.0038, 0.0045]$).

On the opposite corner, subjects commonly considered as requiring procedural and applied skills (Engineering, Economics, and Computer Science) cluster in the lower left corner with negative effects ($\theta \in [-0.09, -0.14]$, $\delta \in [-0.0030, -0.0037]$).

This subject-level pattern reinforces the KC type findings: students typically have higher initial knowledge for declarative Knowledge Components compared to procedural.

\section{Limitations}

Given this work relies on data collected from an in-the-field study tool and not a closely controlled experimental setting, certain biases and limitations exist.

\paragraph{Selection Biases}
\begin{tight_enumerate}
    \item Self-selection: Users demonstrated high self-motivation, limiting generalizability to compulsory classroom populations.
    \item Attrition: Significant dropout and minimum usage thresholds bias the sample toward high-intent learners with an affinity for mobile interfaces.
    \item Interaction bias: The ability to skip exercises allows users to bypass content perceived as redundant, too easy, or too difficult, potentially skewing observed correctness rates.
\end{tight_enumerate}

\paragraph{AI-Generated Content}
\begin{tight_enumerate}
    \item Lack of oversight: Unsupervised LLM generation of KCs and exercises may introduce systematic biases or quality variations absent in expert-curated content.
    \item Classification errors: Automated tagging of course levels, subjects, and KC types limits the reliability of conclusions drawn from the ablation study.
    \item Curriculum alignment: KCs are generated from user-uploaded materials rather than standardized curricula, limiting cross-student comparisons and alignment to external standards.
\end{tight_enumerate}

\paragraph{Student-KC Confounding}
Because KCs are generated uniquely for each student's materials, most KCs are practiced by only one student, making it difficult to fully separate student effects from KC effects. For students practicing few KCs, the estimated learning rate reflects both student ability and KC-specific difficulty. However, the ablation study shows that controlling for course subject, level, and KC type does not reduce the consistency of student learning rates (the IQR of opportunities to reach 80\% mastery remains [7.36, 8.63] in the extended model, compared to [7.01, 8.25] in the base model). It's possible that LLM-generated KCs exhibit more consistent granularity than human-authored cognitive models, potentially contributing to the observed uniformity.

\paragraph{Outcome Measurement}
This study measures short-term mastery within the platform; long-term retention and transfer to external assessments were not evaluated.

\section{Conclusion}

\subsection{Summary of Results}
This study demonstrates that a fully automated system using LLM-generated Knowledge Components and exercises can support effective learning of well-defined, isolated skills at scale. Students utilizing Campus AI reached 80\% mastery in a median of 7.22 practice opportunities, a figure strikingly comparable to the 6.54 opportunities reported for expert-designed content \cite{koedinger2023astonishing}.

Critically, we successfully replicated the "learning rate regularity" in this novel, automated context. The standard deviation of student learning rates ($\text{SD}[\delta_s] = 0.0122$) closely matches the $0.015$ reported in prior work, while initial knowledge varied substantially more ($\text{IQR} = 15.32\%$ vs.\ $0.26\%$ for learning rates). The persistence of this regularity suggests that our AI-generated KCs are capturing genuine cognitive units rather than noise, effectively passing a "psychometric Turing test" for educational content.

Additionally, our ablation study aligns with the theoretical predictions of the Knowledge-Learning-Instruction framework \cite{koedinger2012kli}. We observed that declarative KCs (facts, concepts) are associated with higher initial knowledge and faster learning rates compared to procedural KCs (skills, rules). This differentiation further validates that the LLM is not merely generating text, but is successfully categorizing distinct types of knowledge that result in distinct learning behaviors.

\subsection{Implications for Scalable Personalized Learning}
The ability to automatically generate valid KCs and exercises removes the primary bottleneck in Intelligent Tutoring Systems: the reliance on human experts for content tagging. This opens the door to highly personalized learning environments where curricula can be generated on-the-fly from source material while maintaining pedagogical rigor.

Future work can leverage these automated KCs to power Knowledge Tracing models \cite{abdelrahman2023survey}, allowing for dynamic remediation of specific gaps. While our current system focuses on active exercises (e.g., multiple choice, fill-in-the-blank), integrating constructive or interactive activities could further enhance outcomes \cite{chi2014icap}. Finally, while our approach isolates KCs for clean measurement, future research should explore using frameworks like 4C/ID \cite{vanmerrienboer2021four} to scaffold the integration of these isolated KCs into complex, multi-step problem-solving tasks.

\bibliographystyle{splncs04}
\bibliography{research_paper_v2_bibliography}

\end{document}